\documentclass[twocolumn,showpacs,preprintnumbers,amsmath,amssymb]{revtex4}
\usepackage{graphicx}
\usepackage{dcolumn}
\usepackage{bm}
\begin{document}
\title{Stabilization of a two-dimensional quantum electron solid\\ in perpendicular magnetic fields}
\author{M.~Yu.~Melnikov, D.~G. Smirnov, and A.~A. Shashkin}
\affiliation{Institute of Solid State Physics, Chernogolovka, Moscow District 142432, Russia}
\author{S.-H. Huang and C.~W. Liu}
\affiliation{Department of Electrical Engineering and Graduate Institute of Electronics Engineering, National Taiwan University, Taipei 106, Taiwan}
\author{S.~V. Kravchenko}
\affiliation{Physics Department, Northeastern University, Boston, Massachusetts 02115, USA}
\begin{abstract}
We find that the double-threshold voltage-current characteristics in the insulating regime in the ultra-clean two-valley two-dimensional electron system in SiGe/Si/SiGe quantum wells are promoted by perpendicular magnetic fields, persisting to an order of magnitude lower voltages and considerably higher electron densities compared to the zero-field case. This observation indicates the perpendicular-magnetic-field stabilization of the quantum electron solid.
\end{abstract}
\maketitle
\setlength{\parskip}{0pt}

The ground state of a two-dimensional (2D) electron or hole system in the strongly interacting limit at low carrier densities has been predicted to be a Wigner crystal \cite{chaplik1972possible}.  According to numerical calculations, the crystallization is expected when the interaction parameter given by the ratio of the Coulomb and Fermi energies, $r_{\text{s}}=g_{\text{v}}/(\pi n_{\text{s}})^{1/2}a_{\text{B}}$, exceeds about 35 \cite{tanatar1989ground} (here $g_{\text{v}}=2$ is the valley degeneracy, $n_{\text{s}}$ is the areal density of electrons, and $a_{\text{B}}$ is the effective Bohr radius in semiconductor).  It has been theoretically predicted (see, \textit{e.g.}, Refs.~\cite{lozovik1975crystallization,ulinich1978phase,fukuyama1975two,eguiluz1983two}) that the application of a magnetic field perpendicular to the 2D plane should aid in the formation of the crystalline state.  This happens because a perpendicular magnetic field decreases the amplitude of the zero-point vibrations of the carriers at the lattice sites, thereby ensuring the lattice stability.  In particular, the Wigner crystal was predicted to form in a single-valley 2D electron system at Landau filling factors below approximately 0.15 \cite{lam1984liquid,levesque1984crystallization}.  The presence of a residual disorder is expected to lead to an increase in the electron density of the crystallization, at least in zero magnetic field (see, \textit{e.g}., Ref.~\cite{chui1995impurity}).  There have been claims of the observation of a magnetically-induced Wigner crystal on semiconductor surfaces, which were based on the observation of reentrant insulating phases and single-threshold current-voltage ($I$-$V$) characteristics in the insulating phases around the integer and fractional quantum Hall states (see, \textit{e.g.},
Refs.~\cite{andrei1988observation,williams1991conduction,diorio1992reentrant,qiu2012connecting,knighton2014reentrant,qiu2018new,knighton2018evidence,hossain2022anisotropic,falson2022competing,madathil2023moving}),
attributed to the depinning of the Wigner crystal.  However, alternative mundane interpretations of the data were discussed, such as Efros-Shklovskii variable range hopping in strong electric fields \cite{marianer1992effective} or percolation \cite{jiang1991magnetotransport,dolgopolov1992metal,shashkin1994insulating}. It is worth noting that the formation of a 2D Wigner crystal was assumed in interpreting measurements of the changes in the chemical potential with density variation in the integer quantum Hall regime \cite{zhang2014signatures} and measurements of the nuclear magnetic resonance in the integer and fractional quantum Hall regimes \cite{tiemann2014nmr}.  On the other hand, the observation of a magnetically-induced triangular lattice Wigner crystal was reported in Ref.~\cite{tsui2024direct} using scanning tunneling microscopy in bilayer graphene near the fractional quantum Hall states where the electron system at higher fillings should be in the metallic regime.

Recently, two-threshold $V$-$I$ characteristics that reveal the signature of a quantum Wigner solid and exclude mundane interpretations in terms of percolation or overheating have been observed in high-mobility silicon metal-oxide-semiconductor field-effect transistors \cite{brussarski2018transport} and SiGe/Si/SiGe quantum wells \cite{melnikov2024collective} in a zero magnetic field.  The $V$-$I$ characteristics are strikingly similar to those observed for the collective depinning of the vortex lattice in type-II superconductors \cite{blatter1994vortices} with the voltage and current axes interchanged. The results can be described by a phenomenological theory of the collective depinning of elastic structures, which corresponds to the thermally activated transport accompanied by a peak of generated broadband current noise between the dynamic and static thresholds; the solid slides with friction as a whole over a pinning barrier above the static threshold.

The high uniformity of samples is crucial for observing the double-threshold voltage-current characteristics, and the findings should be universal for other 2D electron systems (see Ref.~\cite{melnikov2024collective}).  In high-mobility Si MOSFETs, the disorder has a short-range character caused by the residual point scatterers at the interface and the low-temperature mobility reaches approximately 3~m$^2$/Vs, whereas in a class of the purest accessible semiconductor heterostructures (including electron and hole GaAs/AlGaAs heterostructures; ZnO/MgZnO heterostructures; AlAs quantum wells; SiGe/Si/SiGe quantum wells), the long-range disorder potential due to background impurities is present and the carrier mobility is higher by orders of magnitude.  As inferred from both the level and character of the disorder potential, Si MOSFETs and unprecedentedly high-mobility heterostructures belong to different classes of electron systems.  Similar results obtained in Si MOSFETs and SiGe/Si/SiGe quantum wells show the generality of the effect for different classes of electron systems.

In this Letter, we find that perpendicular magnetic fields promote the double-threshold $V$-$I$ characteristics in the insulating regime in the ultra-clean two-valley two-dimensional electron system in SiGe/Si/SiGe quantum wells.  The double-threshold behavior arises at an order of magnitude lower voltages and considerably higher electron densities compared to the zero-field case so that the corresponding filling factor at high magnetic fields is equal to $\nu\approx0.27$.  This observation indicates the perpendicular-magnetic-field stabilization of the quantum electron solid, which is consistent with theoretical predictions.

Data were obtained on ultra-high mobility SiGe/Si/SiGe quantum wells similar to those described in Refs.~\cite{melnikov2015ultra,melnikov2017unusual,melnikov2024triple}.  The low-temperature electron mobility in these samples reaches $\approx200$~m$^2$/Vs.  The approximately 15~nm wide silicon (001) quantum well is sandwiched between Si$_{0.8}$Ge$_{0.2}$ potential barriers.  The samples were patterned in Hall-bar shapes with the distance between the potential probes of 100~$\mu$m and width of 50~$\mu$m using photo-lithography.  The triple-top-gate design included the main Hall-bar gate, the contact gate, and the depleting gate, separated by SiO as an insulator; this dramatically reduced the contact resistances and suppressed the shunting channel between the contacts outside the Hall bar, which manifested itself at the lowest electron densities in the insulating regime (for more details, see Ref.~\cite{melnikov2024triple}).  No additional doping was used, and the electron density was controlled by applying a positive dc voltage to the gate relative to the contacts.  Measurements were conducted in an Oxford TLM-400 dilution refrigerator equipped with a magnet. The voltage was applied between the source and the nearest potential probe over a distance of 25~$\mu$m.  The current was measured by a current-voltage converter connected to a digital voltmeter.  The voltage-current curves displayed slight asymmetry upon reversal of the voltage.  We plotted the negative part versus the absolute voltage value for ease of representation.  The electron density was determined by analyzing Shubnikov-de~Haas oscillations in the metallic regime using a standard four-terminal lock-in technique.  We applied saturating infrared illumination for several minutes to the samples, after which the quality of contacts enhanced and the electron mobility increased, as had been found empirically \cite{melnikov2015ultra,melnikov2017unusual}.  The contact resistances measured below 10~kOhm.  Experiments were conducted on three samples, and the obtained results were similar.

\begin{figure}
\scalebox{.7}{\includegraphics[width=\columnwidth]{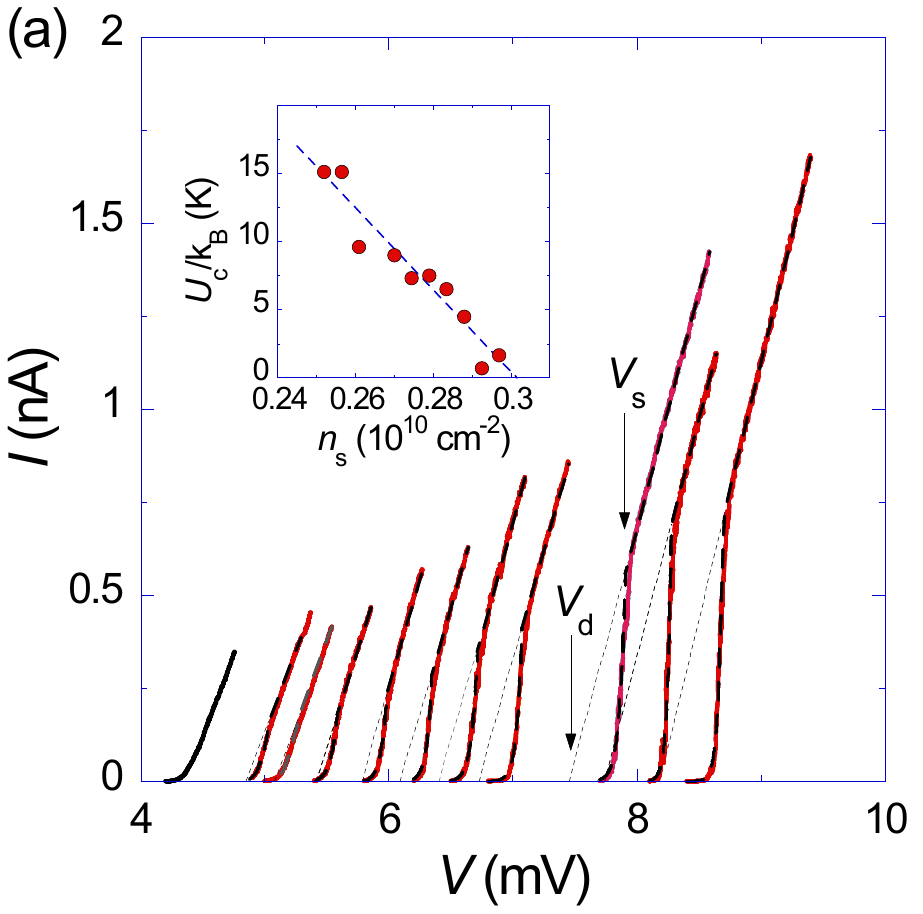}}
\scalebox{.68}{\includegraphics[width=\columnwidth]{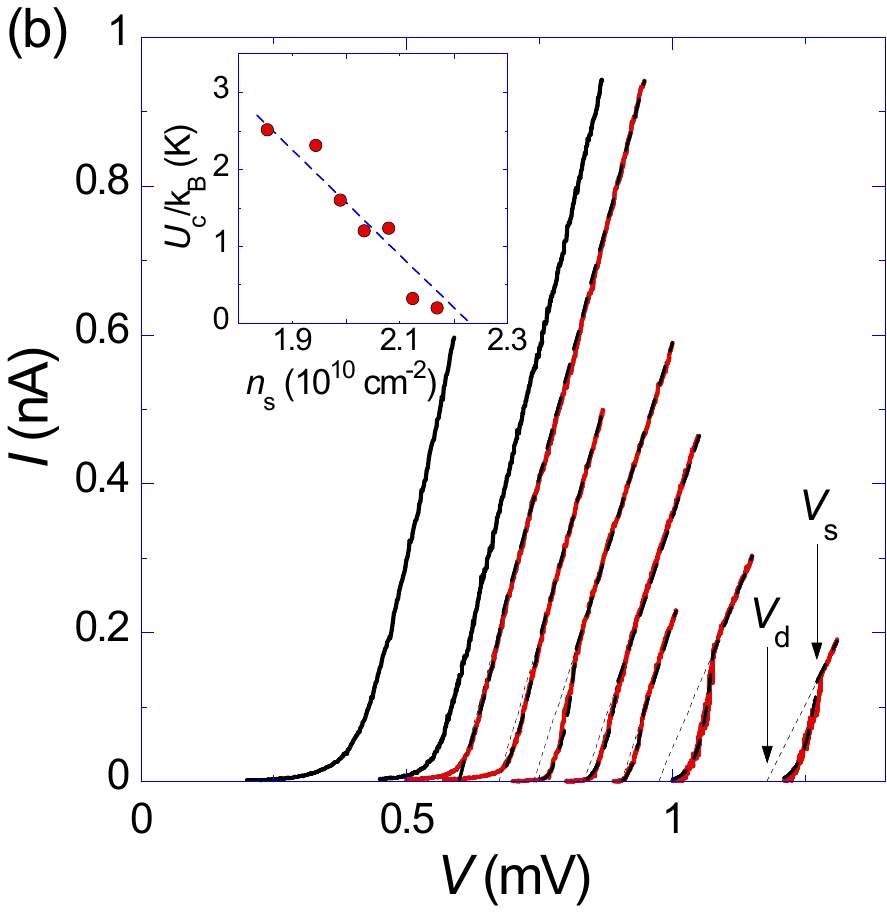}}
\scalebox{.7}{\includegraphics[width=\columnwidth]{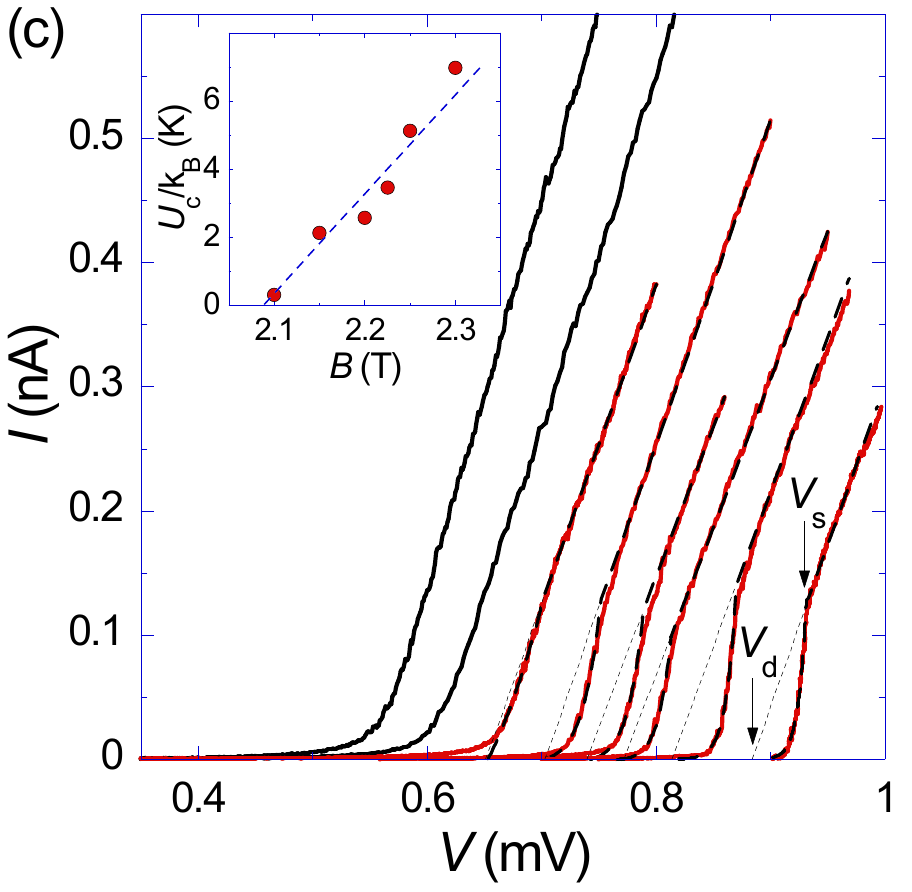}}
\caption{(a) Voltage-current characteristics in zero magnetic field at $T=60$~mK at electron densities (in units of $10^{10}$~cm$^{-2}$, left to right): 0.306, 0.297, 0.293, 0.288, 0.284, 0.279, 0.275, 0.270, 0.261, 0.257, 0.252.  (b) Voltage-current characteristics in $B=3$~T at $T=60$~mK at electron densities (in units of $10^{10}$~cm$^{-2}$, left to right): 2.30, 2.21, 2.17, 2.12, 2.08, 2.03, 1.99, 1.94, 1.85.  (c) Voltage-current characteristics at $T=60$~mK for $n_{\text{s}}=1.67\times 10^{10}$~cm$^{-2}$ in different magnetic fields (left to right): 2, 2.05, 2.1, 2.15, 2.2, 2.225, 2.25, and 2.3~T.  Also shown are the dynamic threshold $V_{\text d}$ obtained by the extrapolation (dotted line) of the linear part of the $V$-$I$ curves to zero current and the static threshold $V_{\text s}$.  The dashed lines are fits to the data, see text.  Insets: activation energy $U_{\text c}$ as a function of the electron density in (a, b) and the magnetic field in (c).  The dashed line is a linear fit.}
\label{fig1}
\end{figure}

\begin{figure}
\scalebox{.7}{\includegraphics[width=\columnwidth]{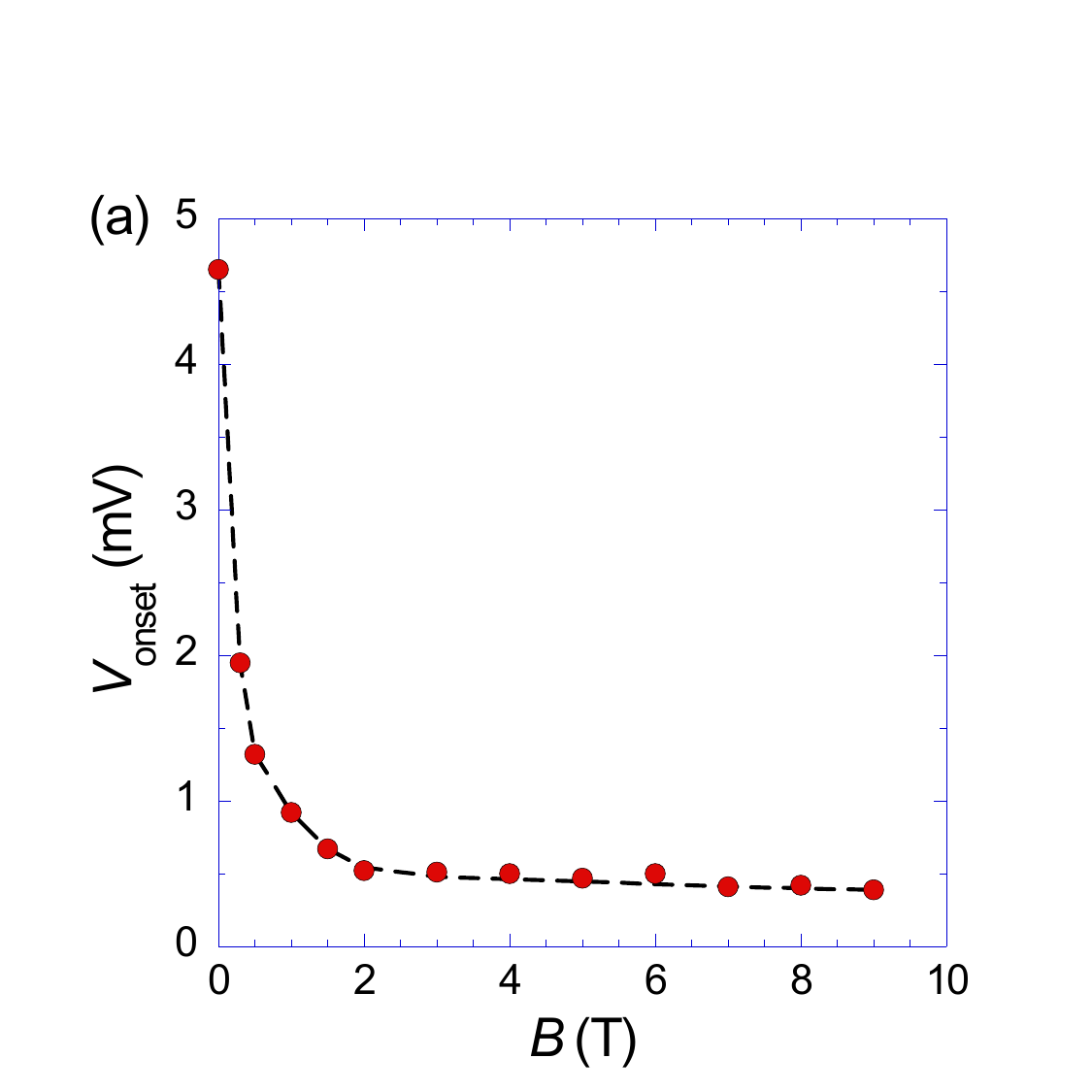}}
\scalebox{.7}{\includegraphics[width=\columnwidth]{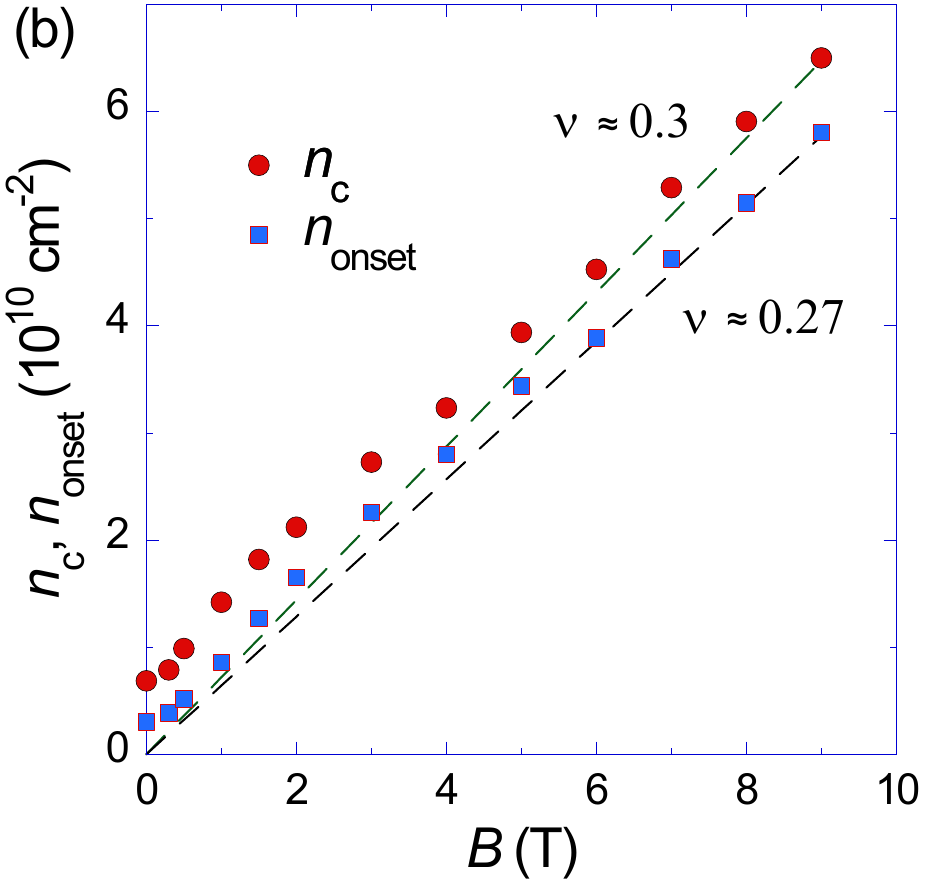}}
\caption{(a) The voltage $V_{\text{onset}}$ for the onset of the double-threshold $V$-$I$ curves at $T=30$~mK as a function of the magnetic field.  The dashed line is a guide to the eye. (b) The corresponding electron density $n_{\text{onset}}$ for the onset of the double-threshold $V$-$I$ curves at $T=30$~mK as a function of the magnetic field (squares).  Also shown is the critical density $n_{\text c}$ for the metal-insulator transition versus magnetic field (circles).  The dashed lines indicate the slopes of the dependences at high $B$.}
\label{fig2}
\end{figure}

In Fig.~\ref{fig1}~(a), we show a set of voltage-current characteristics at a temperature of 60~mK in zero magnetic field at different electron densities in the insulating regime $n_{\text{s}}<n_{\text{c}}$ (here $n_{\text{c}}\approx 0.7\times 10^{10}$~cm$^{-2}$ is the critical density for the metal-insulator transition in our samples); the corresponding interaction parameter $r_{\text{s}}$ exceeds 20 at these values of $n_{\text{s}}$.  Two-threshold voltage-current curves are observed at electron densities below $n_{\text{s}}\approx 0.3\times 10^{10}$~cm$^{-2}$.  With increasing applied voltage, the current stays near zero up to the first threshold voltage.  Then, the current increases sharply until the second threshold voltage is reached, above which the slope of the $V$-$I$ curves is significantly reduced, and the behavior becomes linear although not ohmic, consistent with the previously obtained results \cite{brussarski2018transport,melnikov2024collective}.

The main result of this Letter that the double-threshold voltage-current characteristics are promoted by perpendicular magnetic fields, $B$, is shown in Fig.~\ref{fig1}~(b) at a temperature $T=60$~mK for $B=3$~T.  One can see from the figure that the double-threshold behavior arises at an order of magnitude lower voltages and considerably higher electron densities compared to the zero-field case.  The peak of broadband current noise between the two threshold voltages is observed in the absence or presence of the magnetic field (not shown here).  In this Letter, we focus on the range of electron densities near the onset of the double-threshold $V$-$I$ curves.

Voltage-current characteristics at a temperature $T=60$~mK for $n_{\text{s}}=1.67\times 10^{10}$~cm$^{-2}$ in different magnetic fields are shown in Fig.~\ref{fig1}~(c).  The double-threshold voltage-current curves arise as the magnetic field is increased.

A phenomenological theory of the collective depinning of elastic structures was adapted for an electron solid in Ref.~\cite{brussarski2018transport}.  As the applied voltage increases, the depinning of the electron solid is indicated by the appearance of a current.  Between the dynamic ($V_{\text d}$) and static ($V_{\text s}$) thresholds, the collective pinning of the electron solid occurs, and the transport is thermally activated: $I=\sigma_0\,(V-V_{\rm d})\,\exp[-U_{\rm c}(1-V/V_{\rm s})/k_{\rm B}T]$, where $U_{\rm c}$ is the maximal activation energy of the pinning centers, $\sigma_0$ is a coefficient, and $V_{\rm d}$ corresponds to the pinning force.  When the voltage exceeds the static threshold, the electron solid slides with friction over a pinning barrier, as determined by the balance of the electric, pinning, and friction forces, resulting in linear $V$-$I$ characteristics: $I=\sigma_0\,(V-V_{\rm d})$.  The corresponding fits, shown by the dashed lines in Fig.~\ref{fig1}~(a, b), describe well the experimental two-threshold $V$-$I$ characteristics in the absence or presence of a magnetic field; note that near the onset, the double-threshold $V$-$I$ curves are not too sharp to be described by the theoretical fits over some interval of voltages below $V_{\text s}$.  The fact that the coefficient $\sigma_0$ does not change much with respect to $n_{\text s}$ and $B$ indicates that the solid motion with friction is controlled by weak pinning centers \cite{blatter1994vortices}.  The activation energy $U_{\text c}$ obtained from the fits is plotted as a function of the electron density in the insets to Fig.~\ref{fig1}~(a, b, c).  The electron density, obtained by a linear extrapolation of the $U_{\text c}(n_{\text s})$ dependence to zero, corresponds to the onset density of the two-threshold $V$-$I$ curves at the lowest temperatures.  Near the onset of the double-threshold $V$-$I$ curves, the behavior of $U_{\text c}$ with increasing magnetic field (the inset to Fig.~\ref{fig1}~(c)) is similar to that of $U_{\text c}$ with decreasing electron density.

In Fig.~\ref{fig2}~(a), we plot the voltage $V_{\text{onset}}$, determined by the linear extrapolation to zero current of the $V$-$I$ curve at the onset of the double-threshold behavior at $T=30$~mK, as a function of the magnetic field.  The value $V_{\text{onset}}$ drops by an order of magnitude with increasing magnetic field up to $B\approx 2$~T.  In higher magnetic fields, this continues to decrease with increasing $B$ at much slower rate.

\begin{figure}[t]
\scalebox{.7}{\includegraphics[width=\columnwidth]{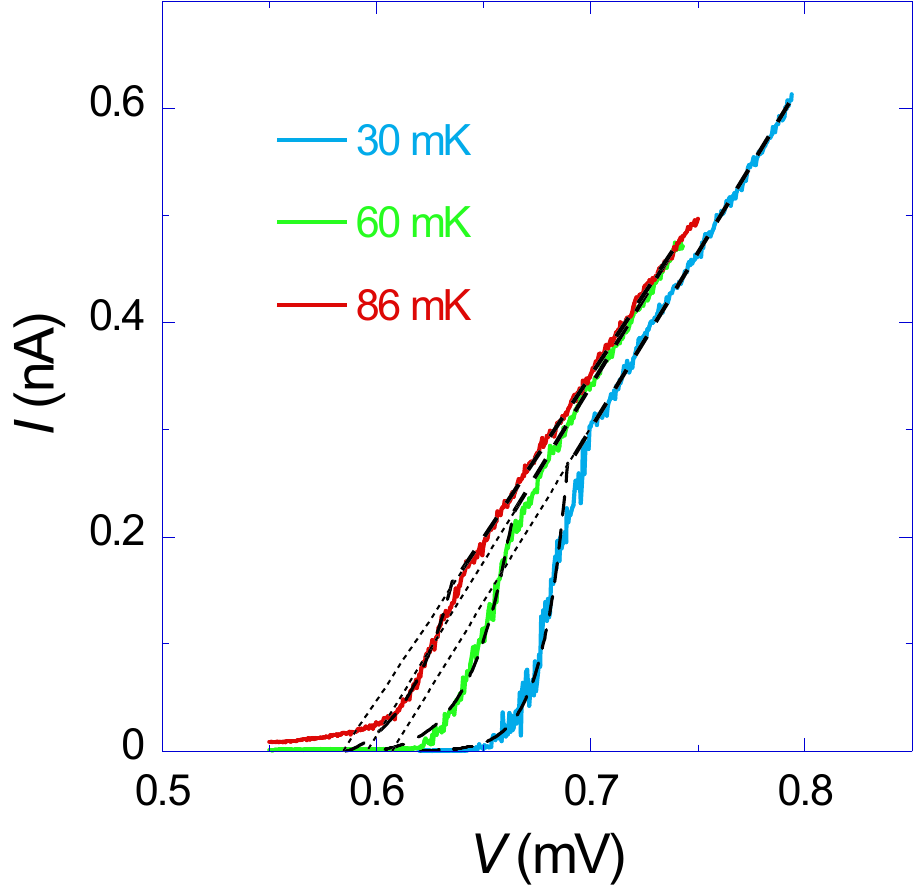}}
\caption{Voltage-current characteristics at $B=4$~T and $n_{\text s}=2.72\times 10^{10}$~cm$^{-2}$ at different temperatures.  The fits corresponding to the activation energy $U_{\text c}=1.5$~K are shown by dashed lines.}
\label{fig3}
\end{figure}

In Fig.~\ref{fig2}~(b), we plot the electron density $n_{\text{onset}}$ for the onset of the double-threshold $V$-$I$ curves at $T=30$~mK as a function of the magnetic field (squares).  Also shown is the critical density $n_{\text c}$ for the metal-insulator transition versus magnetic field (circles), determined by vanishing nonlinearity of the $V$-$I$ curves when extrapolated from the insulating phase, as described in Ref.~\cite{melnikov2019quantum}.  Both values increase with the magnetic field so that for all magnetic fields, $n_{\text{onset}}$ lies below $n_{\text c}$.  Both dependences tend at high $B$ to linear dependences that correspond to filling factor $\nu=n_{\text s}hc/eB\approx 0.27$ for $n_{\text{onset}}$ and $\nu\approx 0.3$ for $n_{\text c}$.

In Fig.~\ref{fig3}, we show the $V$-$I$ characteristics at $B=4$~T and $n_{\text s}=2.72\times 10^{10}$~cm$^{-2}$ at different temperatures.  The fits corresponding to the activation energy $U_{\text c}=1.5$~K describe the data well.  Unlike the $B=0$ case, where the $V$-$I$ curves saturate below $T=60$~mK \cite{melnikov2024collective}, the $V$-$I$ curves in a strong magnetic field follow the expected temperature dependence down to $T=30$~mK, ensuring that there is no overheating down to the lowest temperatures used in our experiments.

The observed increase of the onset density $n_{\text{onset}}$ for the double-threshold $V$-$I$ curves with increasing magnetic field indicates the stabilization of the quantum electron solid in perpendicular magnetic fields.  This is consistent with the theoretical prediction that the application of a perpendicular magnetic field should promote the formation of the Wigner solid by decreasing the amplitude of the zero-point vibrations of the electrons at the lattice sites \cite{lozovik1975crystallization,ulinich1978phase,fukuyama1975two,eguiluz1983two}.  The corresponding filling factor $\nu\approx 0.27$ observed in our two-valley 2D electron system is in reasonable agreement with the theoretical prediction that in a disorderless single-valley 2D electron system, the Wigner crystal should form at filling factors $\nu\lesssim0.15$ \cite{lam1984liquid,levesque1984crystallization}.  It is likely that the intermediate insulating region between $n_{\text{onset}}(B)$ and $n_{\text c}(B)$ that precedes the electron solid can be attributed to a fluctuation region.  The observed drop of the onset voltage $V_{\text{onset}}$ for the double-threshold $V$-$I$ curves with increasing magnetic field also reflects the stabilization effect of perpendicular magnetic fields.  One can give a qualitative account of the behavior of $V_{\text{onset}}$.  In zero magnetic field, the amplitude of the zero-point vibrations of the electrons at the lattice sites is relatively large, and the electron solid is easy to deform near the pinning centers.  As a result, the electrons should be pinned strongly by the pinning centers, which corresponds to a relatively large onset voltage.  In strong perpendicular magnetic fields, the opposite is the case.  The high rigidity/uniformity of the electron solid in strong perpendicular magnetic fields may also be responsible for the observed strong temperature dependence of the $V$-$I$ curves down to the lowest temperatures, unlike the $B=0$ case.

The reentrant behavior for the critical density $n_{\text c}$ for the metal-insulator transition in SiGe/Si/SiGe quantum wells, manifested as a minimum on the metal-insulator phase boundary in a magnetic field $B\approx 0.3$~T at $\nu=1$, was observed using the cut-off resistivity criterion \cite{melnikov2019quantum}. As seen from Fig.~\ref{fig2}~(b), this behavior turns out to be less pronounced when using the criterion of vanishing nonlinearity of the $V$-$I$ curves; namely, the data points for both $n_{\text c}$ and $n_{\text{onset}}$ in $B=0.3$~T deviate downward, reflecting the expected behavior.

In summary, we have found that perpendicular magnetic fields promote the double-threshold $V$-$I$ characteristics in the insulating regime in the ultra-clean two-valley two-dimensional electron system in SiGe/Si/SiGe quantum wells.  The double-threshold behavior arises at an order of magnitude lower voltages and considerably higher electron densities compared to the zero-field case so that the corresponding filling factor at high magnetic fields is equal to $\nu\approx0.27$.  This observation indicates the perpendicular-magnetic-field stabilization of the quantum electron solid, which is consistent with theoretical predictions.  Further in-depth theoretical consideration of the effects is needed.

The ISSP group was supported by the RF State Task.  The NTU group acknowledges support by the Ministry of Science and Technology, Taiwan (Project No.\ NSTC 113-2634-F-A49-008).  S.V.K. was supported by NSF Grant No.\ 1904024.


\end{document}